\documentclass{icrc}

\usepackage{times}
\usepackage{graphicx} 
\begin{document}

\title{The AIRES system for air shower simulations. An update.}
\author[1]{S. J. Sciutto$^{(a)}$}
\affil[1]{Departamento de F\'{\i}sica, Universidad Nacional de La
Plata, Argentina}

\correspondence{S. J. Sciutto (sciutto@fisica.unlp.edu.ar)\\
$^{(a)}$ Fellow of CONICET, Argentina.}

\firstpage{1}
\pubyear{2001}


\maketitle

\begin{table*}[t]
\def\rbeg{\vrule height 9pt depth 0pt width 0pt\relax}
\def\rend{\vrule height 0pt depth 6pt width 0pt\relax}
\begin{center}
\begin{tabular}{|l|p{11.5cm}|}
\hline
\multicolumn{2}{|c|}{{\large\bf
$\vphantom{|^{|^|}_{|_|}}$MAIN CHARACTERISTICS OF AIRES}}\\
\hline
{\bf Propagated particles}
&\rbeg
Gammas. Leptons: $e^{\pm}$, $\mu^{\pm}$.\par
Mesons: $\pi^0$, $\pi^{\pm}$; $\eta$, $K^0_{L,S}$, $K^{\pm}$.
Baryons: $p$, $\bar p$, $n$, $\bar n$, $\Lambda$.\par
Nuclei up to $Z=56$.\par
Neutrinos are generated (in decays) and accounted for their
number and energy, but not propagated.
\rend\\ \hline

{\bf Primary particles}
&\rbeg
All propagated particles can be injected as primary
particles.\par
Multiple and/or ``exotic'' primaries can be
injected using the {\em special primary\/} feature (see text).
\rend\\ \hline

{\bf Primary energy range}
&\rbeg
From 800 MeV to 1 ZeV ($10^{21}$ eV).
\rend\\ \hline

{\bf Geometry and environment}
&\rbeg
Incidence angles from vertical to horizontal showers.\par
The Earth's curvature is taken into account for all inclinations.\par
Realistic atmosphere  (Linsley model).\par
Geomagnetic deflections: The geomagnetic
field can be calculated using the IGRF model \citep{sjs99}.
\rend\\ \hline

{\bfseries Propagation (general)}
&\rbeg
Medium energy losses (ionization).\par
Scattering of all charged particles including corrections for finite
nuclear size.\par
Geomagnetic deflections.
\rend\\ \hline

{\bfseries Propagation: {\itshape Electrons and gammas}}
&\rbeg
Photoelectric and Compton effects.\par
Bremsstrahlung and $e^+e^-$ pair production.\par
Emission of knock-on electrons.\par
Positron annihilation.\par
LPM effect, and dielectric suppression.\par
Photonuclear reactions.
\rend\\ \hline

{\bfseries Propagation: {\itshape Muons}}
&\rbeg
Bremsstrahlung and muonic pair production.\par
Emission of knock-on electrons.\par
Decay.
\rend\\ \hline

{\bfseries Propagation: {\itshape Hadrons and nuclei}}
&\rbeg
Hadronic collisions using the EHSA (low energy) and QGSJET or SIBYLL (high
energy).\par
Hadronic cross sections are evaluated from fits to
experimental data (low energy), or to QGSJET or SIBYLL predictions
(high energy).\par
Emission of knock-on electrons.\par
Decay of unstable hadrons.
\rend\\ \hline

{\bfseries Statistical sampling}
&\rbeg
Particles are sampled by means of the Hillas thinning algorithm
\citep{h81}, extended to allow control of maximum weights.
\rend\\ \hline

{\bfseries Main observables}
&\rbeg
Longitudinal development of all particles recorded in up to 510
observing levels.\par
Energy deposited in the atmosphere.\par
Lateral, energy and time distributions at ground level.\par
Detailed list of particles reaching ground, and/or crossing
predetermined observing levels.
\rend\\ \hline

\end{tabular}
\end{center}
\caption{Main characteristics of the AIRES
air shower simulation system.\label{TAB:features}}
\end{table*}

\begin{abstract}
 A report on the characteristics of ultra-high energy air showers
 simulated with the current version of the AIRES program is
 presented. The AIRES system includes a fast simulating program,
 originally designed on the basis of the well-known MOCCA program, and
 progressively improved and tested. The AIRES algorithms are briefly
 described and some results coming from the simulations are
 analyzed.
\end{abstract}

\section{Introduction}

When an ultra high energy astroparticle interacts with an atom of the
Earth's atmosphere, it produces a shower of secondary particles that
continue interacting and generating mo\-re secondary particles that can
eventually hit the Earth's surface. The study of the characteristics
of such air showers initiated by ultra high energy cosmic rays is of
central importance. This is due to the fact that presently such
primary particles cannot be detected directly; instead, they must be
studied from different measurements of the air showers they produce.

Due to the complexity of the processes that take place during the
development of an air shower, detailed studies of its characteristics
are commonly made with the help of numerical simulations. The
simulating algorithms must take into account all the processes that
significantly affect the behavior of the shower. This includes
electrodynamic interactions, hadronic collisions, photonuclear
processes, particle decays, scattering, etc.

Among all those processes, the hadronic and photonuclear reactions
are, at present, the less understood. In the case of air showers
initiated by ultra-high energy astroparticles ($E\ge 10^{19}$ eV), the
primary particles have energies that are several orders of magnitude
larger than the maximum energies attainable in experimental
colliders. This means that the models used to rule the behavior of
such energetic particles must necessarily make extrapolations from the
data available at much lower energies, and there is still no
definitive agreement about what is the most convenient model to accept
among the several available ones. On the other hand, the results
obtained for the most common shower observables prove to have a non
negligible dependence with the model used to simulate the low energy
hadronic collisions (with energy less than a few hundred GeV). Despite
the fact that both high and low energy hadronic collision models are
tuned using experimental data (energies less than $\approx$ 100 GeV),
their predictions in the transition region do not agree completely,
being this an undesirable characteristic of such models that is still
not completely understood.

The {\bf AIRES system}%
\footnote{{\bf AIRES} is an acronym for
$\underline{\hbox{\bf AIR}}$-shower
$\underline{\hbox{\bf E}}$xtended
$\underline{\hbox{\bf S}}$imulations.}
\citep{sjs99} is a set of programs to simulate air showers. One of
the basic objectives considered during the development of the software is
that of designing the program modularly, in order to make it easier to
switch among the different models that are available, without having to
get attached to a particular one.
The MOCCA code created by A. M. Hillas \citep{h97}
--successfully used to interpret experimental data coming from the
Haverah Park experiment--
has been extensively used as the primary reference when
developing the first version of AIRES \citep{sjs97}, launched four
years ago.

\begin{figure}[ht]
\includegraphics[width=8.3cm]{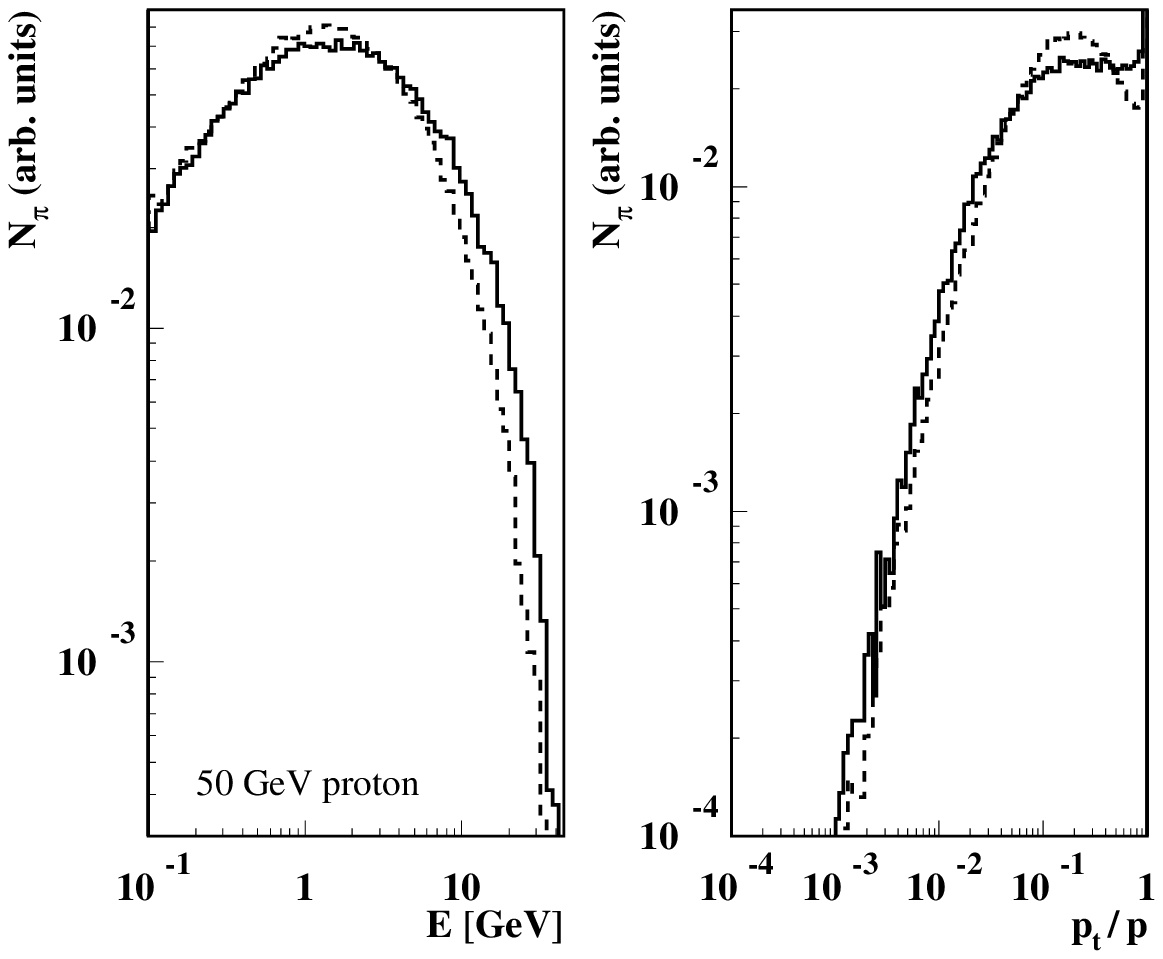}
\caption{Energy and transverse momentum distribution of secondaries
generated by the extended Hillas algorithm (solid lines) and QGSJET
(dashed lines) for collisions generated by 50 GeV
protons.\label{FIG:SQC50gev}}
\end{figure}

\begin{figure}[ht]
\includegraphics[width=8.3cm]{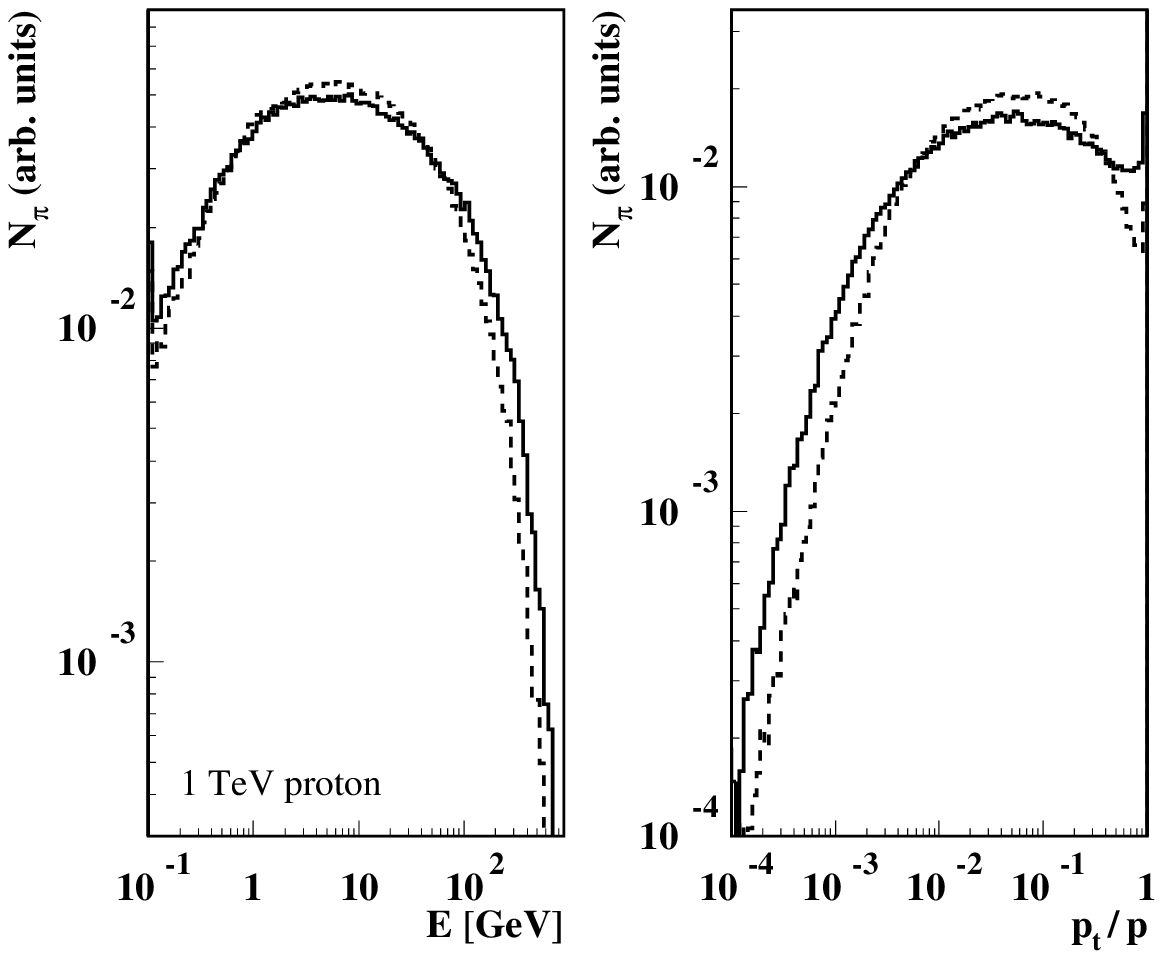}
\caption{Same as figure \ref{FIG:SQC50gev}, but for collisions
generated by 1 TeV protons.\label{FIG:SQC1tev}}
\end{figure}

Later releases have progressively incorporated new developments,
including a complete revision, correction and extension of all the
physical algorithms.

AIRES has been successfully used to study several characteristics of
the showers, including comparison between ha\-dro\-nic models at the
highest energies \citep{doqui99}, influence of the LPM effect
\citep{LPM}, muon bremsstrahlung \citep{mubrem}, and geomagnetic
deflections \citep{GF}. It has also been employed to study the
expected efficiency of the Auger Observatory for detecting
quasi-horizontal showers generated by $\tau$ neutrinos
\citep{Pierretau}, and to calibrate a reconstruction technique for
analyzing quasi-horizontal showers measured at the Haverah Park
experiment \citep{Ave}.

In all the studies performed using AIRES, the results obtained present
a good agreement with experimental data or simulated data coming
from other sources, in particular, a recent comparative study of the
AIRES and CORSIKA programs \citep{cc} shows that the
output from both codes are in excellent agreement.

The aim of this paper is to describe the relevant characteristics
of the AIRES program, stressing on the most recent developments, and
discuss briefly the strategy for future developments.

\section{Characteristics of AIRES}

The AIRES simulation system provides a comfortable environment for
performing realistic simulations taking advantage of present day
computer technology.

Table \ref{TAB:features} summarizes the main characteristics of the
particle propagating engine of AIRES.

The set of particles that are fully propagated by AIRES include the
most commonly observed ones, together with other less numerous but
capable of producing indirectly a non negligible impact on the final
shower observables. All these particles can be injected as shower
primaries. It is also possible to initiate showers produced by
``special'' primaries. This useful feature of AIRES, explained in
detail elsewhere \citep{sjs99} allows to dynamically call a
user-defined module that generates the primary (or primaries) that
starts the shower, thus extending the set of primary particles beyond
the standard ones that are recognized by the propagating engine.

 The interactions indicated in table \ref{TAB:features} represent, for
the case of air showers, the most important ones from the
probabilistic point of view. Particle decays and electromagnetic
interactions are simulated using built-in procedures that have been
developed on the basis of tested and commonly accepted theoretical
formulations. In the particular cases of the LPM effect \citep{LPM}
and the muon bremsstrahlung \citep{mubrem}, the development of such
procedures has included exhaustive studies of the underlying
physics. Additionally, we have recently finished a review
of all the electromagnetic procedures implemented in AIRES, including
the scattering algorithms. After this review, a series of minor
changes were implemented to improve the accuracy and performance of
the program.

However, our knowledge of the hadronic interactions is by far more
incomplete than in the case of the electromagnetic ones. To process
such interactions, it is necessary to rely on a given model, that is
always based on phenomenology. Additionally, in an air shower, the
energy spectrum of the hadrons undergoing inelastic collisions spans
regions where there are no experimental data available, where the only
alternative is to use the extrapolations provided by the available
models.

In AIRES the hadronic collisions are processed by means of two
models, depending on the energy of the projectile: For collisions with
energy less than $\approx$ 100 GeV, an extension of the Hillas Splitting
Algorithm (EHSA) (Hillas, 1981) is used, while for higher energies it is
possible to select between QGSJET \citep{QGSJET} and SIBYLL \citep{SIBYLL}.

In a previous work \citep{doqui99} we made a comparative study of QGSJET and
SIBYLL at the highest primary energy. More recently, and related with
a comparative study AIRES-CORSIKA \citep{cc}, we analyzed the
characteristics of the EHSA. In the following section we summarize
some results of this study.

\section{The low energy hadronic model}

At energies below $\approx$ 100
GeV the high-energy models start to get problems, since particle
production is constrained by the small amount of energy available, and
a low energy mo\-del is necessary to process such interactions.

The low-energy model is of great importance, 
since all signals measured in an EAS experiment are
produced by low-energy particles that come from low-energy
interactions. Especially particle ratios and energies can be
altered by those interactions.

As mentioned, AIRES uses the EHSA in which the initial energy is split
at random into smaller and smaller portions. There are two free
parameters, one regulates the mean energy fraction at which the
splitting occurs and the other controls the number of subsequent
splittings that are applied.  Finally the energy portions are
attributed to pions and nucleons. The EHSA can be easily configured to
approximately emulate the multiplicities and energy distributions of
other models. Cross-sections, transverse momenta
distributions and composition of secondaries need to be inserted from
outside\footnote{The low energy cross sections can easily be obtained
from fits to experimental data.}.

Despite its simplicity, the EHSA can emulate with acceptable quality
the main characteristics of the secondaries produced by other, more
involved, models.

To illustrate this point we show here some results obtained from a
comparison between QGSJET and the EHSA. In figures \ref{FIG:SQC50gev}
and \ref{FIG:SQC1tev} the energy and transverse momentum distribution
of secondary pions are displayed for the cases of 50 GeV and 1 TeV
proton primaries, respectively. The solid (dashed) lines correspond to
the EHSA (QGSJET). The agreement between distributions is acceptable,
and this also applies to other observables and intermediate energies
(not displayed here).
\begin{figure}[t]
\includegraphics[width=8.3cm]{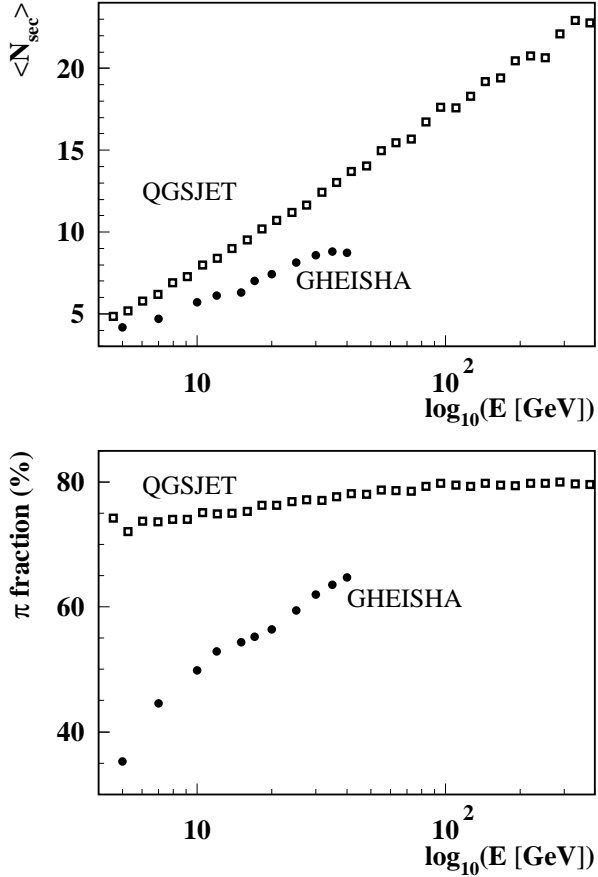}
\caption{Average number of secondaries (top) and fraction of pions
(bottom) corresponding to inelastic collisions initiated by protons
and processed using GHEISHA and QGSJET.\label{FIG:QG}}
\end{figure}

The parameters of the EHSA can be also easily adjusted to match the
distributions coming from other models, including the ones designed to
work at low energies like the well-known GHEISHA model
\citep{GHEISHA}. It is relevant to mention that at the transition
energy ($\approx 100$ GeV) the predictions of GHEISHA and QGSJET do not
agree completely, as it can clearly be seen from figure \ref{FIG:QG},
where the total number of secondaries and fraction of pions, as measured
from proton initiated collisions, are plotted versus the primary
energy. Both plots show an evident discontinuity at the transition
energy that in this case was set to 50 GeV.

We have taken profit of the flexibility of the EHSA to simulate the
behavior of any of the two models to perform a survey of the
influence of the low energy hadronic model on shower observables. From
this study, we have found that there is a moderate dependence of some
non-electromagnetic shower observables like the lateral distribution
of muons, that present variations of up to 40 \% at large distances to
the core, that are clearly dependent on the settings of the low energy
hadronic model. A more detailed study on this subject is in progress
and will be reported elsewhere.

\section{Final remarks}

We have presented some of the main features of the AIRES system for
air shower simulations, stressing on the recent developments than can
be summarized as follows:
\begin{itemize}
\item All the electromagnetic shower algorithms have been checked.
\item Complete treatment of muon radiative processes.
\item Revised extended Hillas Splitting algorithms, tuned to match
data coming from more involved models like GHE\-I\-SHA and QGSJET.
\item Experimental cross sections at low energies.
\item Other minor changes.
\end{itemize}

We are presently working on the release of a new public version of
AIRES that will contain all the reported features. The current status
of the AIRES system can always be checked at the AIRES Web page:\\
{\ttfamily\bfseries www.fisica.unlp.edu.ar/auger/aires}.
\section*{Acknowledgments}
This work was partially supported by Fundaci\'on Antorchas
of Argentina.

\end{document}